# How Anomalous Resistivity Accelerates Magnetic Reconnection

H. Che

*University of Maryland, College Park, MD, 20742, USA  and*

*NASA Goddard Space Flight Center, Greenbelt, MD, 20771, USA*

(Dated: 11 August 2017)

Whether turbulence induced anomalous resistivity (AR) can facilitate a fast magnetic reconnection in collisionless plasma is a subject of active debate for decades. Recent space observations suggest that the reconnection rate can be higher than the Hall-reconnection rate and turbulent dissipation is required. In this paper, using particle-in-cell simulations, we present a case study of how AR produced by Buneman instability accelerates magnetic reconnection. We first show that the AR/drag produced by Buneman instability in a thin electron current layer 1) can dissipate magnetic energy stored in the current layer through dissipation of the kinetic energy of electron beams; 2) The inhomogeneous drag caused by wave couplings spontaneously breaks the magnetic field lines and causes impulsive fast non-Hall magnetic reconnection on electron-scales with a mean rate reaching 0.6 $V_A$. We then show that a Buneman instability driven by intense electron beams around the x-point in a 3D magnetic reconnection significantly enhances the dissipation of the magnetic energy. Electron-scale magnetic reconnections driven by the inhomogeneous drag around the x-line enhances the reconnection electric field and the in-plane perpendicular magnetic field. About 40% of the released magnetic energy is converted into the electron thermal energy by AR while 50% is converted into the kinetic energy of the electron beams through the acceleration by the reconnection electric field. The enhanced magnetic energy dissipation is balanced by a net Poynting flux in-flow. About 10% of the released magnetic energy is brought out by an enhanced Poynting flux out-flow. These results suggest that AR with sufficient intensity and electron-scale inhomogeneity can significantly accelerate magnetic reconnection.

## I. INTRODUCTION

In the Earth's magnetosphere[1–5] and solar and stellar atmospheres[6–8], plasma heating commonly occurs in current sheets of various scales. Different types of macro and micro-instabilities, including both electrostatic(ES) and electromagnetic(EM) instabilities, can be triggered in the current sheets. Some can lead to the merging and rearranging of oppositely directed magnetic field lines and bursty releasing of magnetic energy – a process known as magnetic reconnection.

In magnetic reconnections, the breaking of magnetic field lines requires the ideal magnetohydrodynamics (MHD) frozen-in condition $\mathbf{E} + \mathbf{U} \times \mathbf{B} = 0$ to be broken. In resistive MHD theory, e.g. the established Sweet-Parker model[9], it is the collisional resistivity $\eta$ that breaks the frozen-in condition, and the Ohm's law assumes the well known form $\mathbf{E} + \mathbf{U} \times \mathbf{B} = \eta \mathbf{j}$. However, the collisional resistivity of space plasmas is usually too low to facilitate a sufficiently fast Sweet-Parker magnetic reconnection to explain the observed magnetic energy release in solar flares and magnetospheric substorms.

In thin current sheets, ions and electrons decouple and two-fluid effects dominate[10,11]. In this case the Ohm's law is replaced by the generalized Ohm's law – the first moment of electron Vlasov equation[12,13]:

$$\mathbf{E} + \frac{1}{c}\mathbf{U}_i \times \mathbf{B} = -\frac{m_e}{e}(\partial_t \mathbf{U}_e + \mathbf{U}_e \nabla \cdot \mathbf{U}_e) + \frac{1}{en_e c}\mathbf{j} \times \mathbf{B} - \frac{1}{en_e}\nabla \cdot \mathbb{P}_e - \eta \mathbf{j}_e, \qquad (1)$$

where we used $\mathbf{U}_e \times \mathbf{B}/c \equiv (\mathbf{U}_i - \mathbf{j}/en_e) \times \mathbf{B}/c$ and for simplicity assumed the plasma being fully ionized hydrogen with $n_e = n_i$ where $n_i$ and $n_e$ are ion the electron density respectively. The Hall term[14] is defined as $\mathbf{j} \times \mathbf{B}/c$. Besides the electron Joule heating $\eta \mathbf{j}_e$, the generalized Ohm's law presents more terms that can break the frozen-in condition, including the gradient of electron pressure $\mathbb{P}_e$ and the electron inertia terms (acceleration $\partial_t \mathbf{U}_e$ and convective momentum $\mathbf{U}_e \nabla \cdot \mathbf{U}_e$). The Hall term increases as the spatial scale approaches the ion inertial length $d_i = c/\omega_{pi}$ (where $\omega_{pi} = \sqrt{m_i/4\pi e^2 n_e}$) but decreases as the spatial scale decreases to the electron inertial length $d_e = c/\omega_{pe}$. The Hall term becomes zero at the x-point and thus it can not break magnetic field lines but rather helps to form an x-point configuration through the Hall field[10,15,16]. The inertia and pressure terms are important on electron scales[12,17–19].

Simulations of laminar driven magnetic reconnection, e.g. the GEM magnetic reconnection challenge[12], show that by including Hall-effect reconnection can become dramatically faster than Sweet-Parker reconnection, and its maximum steady reconnection rate can reach 0.1-0.2 $V_A$, where $V_A$ is Alfvén speed. An interesting observation based on a set of particle-in-cell (PIC) simulations is that the maximum Hall-reconnection rate seems universal and is independent of the mechanism that breaks the field lines[20,21]. *Observations and laboratory experiments have discovered the Hall field and Hall current associated with fast collisionless magnetic reconnection*[10,12], *but no direct experimental or observational evidence has been found to show that the rate is irrelevant to the dissipation mechanism*[22–25]. On the other hand, observations of solar flares found that magnetic reconnection may commonly be turbulent as evidenced by the filamentary structure of magnetic field, non-thermal heating observed in X-ray, and coherent radio emissions that are likely results of electron two-stream instability[26–28]. The reconnection rate in solar flares estimated by the variability of magnetic flux around magnetic null-points varies in a wide range, from the moderate 0.1 $V_A$ to very fast 0.5-0.6 $V_A$[29] – much higher than the maximum Hall-reconnection rate from simulations. Note that solar flare magnetic reconnections mostly have strong guide magnetic field. Experiments[30] and simulations[31] show that the reconnection rate of guide field reconnection is lower than the anti-parallel reconnection rate. Therefore the discrepancy between the simulation results and observation is more severe than it appears. New *in situ Magnetospheric Multiscale Science* (MMS) observations of magnetopause reconnection provide the first direct evidence that the reconnection rate is higher than the Hall-reconnection rate and the turbulence induced anomalous resistivity is probably required to explain the enhanced reconnection rate[32]. In addition, Magnetic Reconnection Experiment (MRX) suggests that anomalous resistivity may be able to facilitate a fast Sweet-Parker-like magnetic reconnection[33,34].

It's not surprising that the inconsistency exists between observations and numerical simulations. First, past numerical simulations have focused on a class of driven magnetic reconnections with a limited parameter space[12] which may not be representative of the reconnections in nature; Second, the simulated reconnections can be affected by initializations and boundary condition choices and it is unclear how these affect magnetic reconnection rate and the generation of instabilities and waves on different scales[35]. A recent simulation carried out by Sitnov et al. shows that an internal instability drives a new reconnection in the ion diffusion region (IDR) with an unreported Hall-field pattern[36], while a simulation carried out by Che et al. found that kinetic instabilities driven at the reconnection x-line can significantly increase the reconnection rate[37]. In nature, the ways magnetic reconnection can be triggered are far more diverse than have been explored in simulations, and the process can occur on a wide range of scales simultaneously. More importantly, observations of both magnetospheric magnetic reconnections and solar flares show that magnetic reconnection are generally impulsive and turbulent[22,27,38,39]. While the theory of collisional magnetic reconnection is better established, turbulent magnetic reconnection which involves wave-wave and wave-particle interactions on a broad range of spatial and time scales is still poorly understood.

What is known is that the turbulence effects such as anomalous resistivity (AR) and anomalous viscosity (AV) can dramatically influence the energy dissipation and convective momentum transport. Thus the generalized Ohm's law in Eq. (1) needs to be modified for turbulent reconnection[37]:

$$\langle \mathbf{E} \rangle + \langle \mathbf{U}_e \rangle \times \langle \mathbf{B} \rangle / c = \mathbf{D}_e + \nabla_\perp \cdot \mathbf{\Xi}_e - \frac{m_e}{e}(\partial_t \langle \mathbf{U}_e \rangle + \langle \mathbf{U}_e \partial \cdot \mathbf{U}_e \rangle) - \frac{1}{e\langle n_e \rangle} \partial \cdot \langle \mathbb{P}_e \rangle. \quad (2)$$

where $\mathbf{D}_e \equiv -\langle \delta n_e \delta \mathbf{E} \rangle / \langle n_e \rangle$ is the drag, i.e. AR, $\mathbf{\Xi}_e \equiv -\langle \delta \mathbf{p}_e (\delta \mathbf{U}_e - e \delta \mathbf{A}/m_e c) \rangle / e \langle n_e \rangle$ is AV, $\mathbf{p}_e = n_e \mathbf{U}_e$, and $\mathbf{A}$ is the magnetic vector potential. $\langle ... \rangle$ denotes the ensemble average of the turbulence properties.

AR[40] is induced by ES instabilities, arising from ion-electron drag, and AV[41] is induced by EM instabilities, arising from anomalous momentum transport. Whether AR and AV can facilitate fast magnetic reconnection in collisionless plasma (if so, how it is achieved) is a fundamental problem in both plasma physics and space physics and is a subject of debate for nearly half a century[34,37,42–53]. The problem is particularly difficult because AR and AV are highly nonlinear processes which take part in the even more complex magnetic reconnection processes. Diagnosing the role of AR and AV in magnetic reconnection is a daunting task, and the progress is slow. In observations, since kinetic instabilities generating AR and AV are dominated by electron dynamics, addressing the problem experimentally requires resolution of the electron diffusion region (EDR). Only recently it becomes possible to probe such scales with the launch of the MMS.

The analysis of the generalized Ohm's Law (Eq. 1) using the observational data from the first encounter of a magnetopause reconnection EDR by MMS shows that AR is required to explain the enhanced magnetic reconnection rate[32]. On the other hand, many other MMS observations discovered coherent electrostatic structures and electron heating in reconnection diffusion region, implying that ES instabilities are common in reconnections[32,54–58]. Thus it is necessary for us to revisit how AR affects the process of magnetic reconnection in the EDR, and how AR couples with Hall effect to affect the IDR – both are fundamental problems in plasma physics[59].

In this paper, we concentrate on AR induced by Buneman instability in an unsteady guide-field reconnection. Buneman instability is triggered due to the intense electron beams developed at the x-line in magnetic reconnection and once the electron beam velocity is larger than the electron thermal velocity[40,45,60–63]. Recent MMS observations of magnetospheric reconnection discovered that Buneman instability can efficiently brake the electron jet at the exhaust of the diffusion region[54]. Electron holes driven by Buneman instability are often discovered to associate with electron-scale current sheets in reconnections in the magnetosphere[54,56,64–68]. Intense thin electron current sheet develops near the x-line in the EDR has also been discovered for the first time by MMS[69].

The 3D guide field magnetic reconnection simulation first published by Che et al. (2011)[37] is so far the only study that clearly demonstrates the dramatic enhancement of reconnection rate by turbulence, because in this particular case turbulence effects can be separated from Hall-effect by comparing 3D PIC simulations with a 2D benchmark simulation. In this 3D guide field magnetic reconnection simulation, intense electron beam develops at late stage when the current sheet near the x-line becomes very thin (with width $\sim d_e$) and triggers a Buneman instability. At the late stage of the reconnection, the nonlinear evolution of Buneman instability triggers whistler wave turbulence and then induces anomalous viscosity that maintain the fast reconnection[37]. Careful analysis of this simulation can enable us to gain better insights into how turbulence effects accelerate magnetic reconnections.

The outline of the paper is as follows: In § II A we first describe the simulations used in this analysis; then in § II B we present an analysis of Buneman instability induced AR in a thin current layer, in which we show the inhomogeneous drag can spontaneously break magnetic field lines to produce electron scale magnetic reconnections that are not affected by the Hall-effect. In § II C we examine the properties of the Buneman instability induced drag in a 3D force-free magnetic reconnection and show how drag helps to break the field lines, accelerate magnetic reconnection, and dissipate magnetic energy. The conclusions and a discussion are presented in § III.

## II. SIMULATIONS AND ANALYSIS

### A. Simulations

Four PIC simulations using the P3D code[70] are employed in this analysis, including simulations of Buneman instability in a thin current layer in 2D and 3D, and magnetic reconnection also in 2D and 3D.

The initialization of these simulations are similar to or based on previous papers[37,51,71]. The differences between the Buneman and reconnection simulations are 1) the width of the initial current sheet in the former is of electron inertial length scale $d_e$ while the latter is of ion inertial length scale $d_i$; and 2) contrary to reconnection simulations, no initial perturbation is applied in the Buneman instability simulation so that the Buneman instability can develop well before magnetic reconnection starts spontaneously.

The parameters of 3D Buneman instability are the same as those in the simulation reported previously[51]. The initial magnetic field has a force-free configuration with $B_{x,0}/B_0 = \tanh[(y - L_y/2)/w_0]$, where $w_0$ and $L_y$ are the half-width of the initial current sheet and the simulation box size in y-direction, respectively. The guide magnetic field $B_g^2 = B_{z,0}^2 = B_0^2 - B_{x,0}^2$ is chosen so that the total magnetic field $|B| \equiv \sqrt{26}B_0$. We choose $w_0 = d_e$ in the Buneman instability simulations and $w_0 = 0.5d_i$ in the magnetic reconnection simulations. Simulations use periodic boundary conditions in the $x$ and $z$ directions, and conducting boundary condition in the $y$ direction. The initial plasma temperature and density are isotropic and uniform, with $T_{e0} = T_{i0} = 0.04 m_i V_{A0}^2$, where $V_{A0} = B_0/(4\pi n_0 m_i)^{1/2}$ is the asymptotic ion Alfvén speed. A mass ratio $m_i/m_e = 100$ is used in all simulations. The force-free condition requires $\mathbf{j}_e \times \mathbf{B} = 0$, thus initially $j_{ex}/j_{ez} = B_x/B_z$.

The domain of 3D simulation of Buneman instability has dimensions $L_x \times L_y \times L_z = 1d_i \times 1d_i \times 2d_i$ with grid number $512 \times 512 \times 1024$. The particle number per cell is 100. The initial electron drift velocity is $v_{de} \sim 9V_{A0} \sim 3v_{tez,0}$, which is large enough to trigger electrostatic Buneman instability. The total simulation time is $\omega_{pe,0}t = 160$ where $\omega_{pe,0} \equiv (4\pi n_0 e^2/m_e)^{1/2}$ is the initial electron plasma frequency. The same initialization is used in the corresponding 2D simulation except that only 1 cell is used in the x-direction. The 2D simulation is in the y-z plane in which the waves of Buneman instability propagate along z and form 2D electron holes in the y-z plane.

The parameters for the 3D magnetic reconnection simulation are the same as those in the simulation published in a previous paper[37]. The domain has dimensions $L_x \times L_y \times L_z = 4d_i \times 2d_i \times 8d_i$ with grid number $512 \times 256 \times 1024$. The particle number per cell is 20. The initial temperature is the same as that in the Buneman instability simulation, but the initial electron drift is $v_{de} \sim 4V_{A0} \sim v_{te,0}$, much smaller than that in the Buneman instability simulation, thus the Buneman instability is much weaker in the 3D reconnection simulation. The total simulation time is $\Omega_{i0}t = 4$, where $\Omega_{i0}$ is the asymptotic ion gyro-frequency and $\omega_{pe,0}/\Omega_{i0} = 200$. The corresponding 2D reconnection domain is $L_x \times L_y = 4d_i \times 2d_i$. The total simulation time $\Omega_{i0}t = 4$ is short enough to prevent the inflow $V_{in}$ and outflow recirculation due to $V_{in} << V_{out} < V_{A0}$ and the passing time of the flow is $<< 4d_i/V_{A0} = 4\Omega_{i0}^{-1}$.

The simulation quantities are dimensionless, with magnetic field normalized to the asymptotic $B_0$, density to the asymptotic $n_0$, and velocity to $V_{A0}$. The units of time and distance are $\Omega_{i0}^{-1}$ (or $\omega_{pe,0}^{-1}$ in some cases) and $d_i$ respectively. In the dimensionless units, electric field $E_0 = V_{A0}B_0/c$.

## B. Buneman Instability in a thin current sheet and the Spontaneous Fast Electron-Scale Magnetic Reconnection

In the 3D simulation of the thin current sheet a Buneman instability is driven along the magnetic field (which is close to the z-direction) at $\omega_{pe,0}t \sim 40$ with the initial growth rate $\sqrt{3}/2(m_e/2m_i)^{1/3}\omega_{pe,0} \sim 0.1\omega_{pe,0}$, consistent with the linear growth rate of Buneman instability in cold plasma. The electric fluctuation $\langle \delta E_z^2 \rangle^{1/2}$ ($\langle ... \rangle$ denotes the average over z, which approximates the ensemble average) generated by the Buneman instability reaches its peak at $\omega_{pe,0}t \sim 70$ and then decays and saturates at $\omega_{pe,0}t \sim 160$. As the magnitude of electric fluctuations become large enough to trap electrons and form electron holes with electric potential satisfying $\phi > kT_e$, the Buneman instability enters the nonlinear stage and the growth rate decreases due to thermal effects[71]. The fast adiabatic energy exchange between the electron holes and the trapped electrons causes rapid phase mixing and heating, leading to the decay of waves and the de-trapping of electrons until the saturation of the Buneman turbulence when the electron holes break up[71]. Previous study[51] shows that the kinetic energy of the current and the associated magnetic energy are dissipated by AR in the form of drag $\mathbf{D}_e$. The dissipated kinetic energy is converted into parallel electron heating while the dissipated magnetic energy is converted into perpendicular electron heating. At the same time, besides the electron holes, an inductive electric field is produced due to the dissipation of magnetic energy.

A natural question is whether AR can break magnetic field lines as the magnetic energy is dissipated. Indeed, after the Buneman instability is triggered at $\omega_{pe,0}t \sim 40$, electron-scale magnetic reconnections appear at different locations along the mid-plane of the current sheet in the simulation. Examples of the magnetic field lines averaged over z at $\omega_{pe,0}t =48, 60$ are shown in Fig. 1. Magnetic field perpendicular to the current sheet $\langle B_y \rangle$, which was originally zero, is generated at $x = 0.7d_i$ and an x-line is formed at $\omega_{pe,0}t =60$. A clear manifestation of topological change of the magnetic fields is shown in the magnetic vector potential $\langle A_z \rangle$ with amplitude varying from 0 to -0.5 $B_0/d_i$ (Fig.2, multimedia view). As a result of the magnetic reconnection, the inductive electric field $\langle E_z \rangle$ is produced with a value as high as 0.6 $V_{A0}B_0/c$. The corresponding localized intense electric field $E_z$ increases to as high as $40E_0$ and the electron density fluctuations can be as high as $\delta n/n_0 \sim 0.8$ in the mid-plane, creating the observed intense drag $\langle \delta n_e \delta E_z \rangle$.

A remarkable feature of the electron-scale magnetic reconnection is that contrary to ion-scale reconnection, contribution from Hall-effect is negligible. In Hall-reconnections the Hall effect generates an out-of-plane *Hall magnetic field*[10,15]. Hall field has a quadrupole structure in anti-parallel reconnections[16], and the quadrupole is distorted in guide-field reconnections[31,72]. Such a quadrupole Hall field is clearly absent in the simulation as shown in the out-of-plane magnetic field $\langle B_z \rangle - B_g$ produced during the Buneman instability (Fig. 1). What we observe is a very small decrease of $B_z$ caused by the dissipation of $j_{ex}$ ($<< j_{ez}$). Hall effect is produced due to the decoupling of ions and electrons. The reason for the disappearance of Hall effect is that the electrons are fully demagnetized inside the electron holes and the wave-electron interactions on electron-scale dominate the dynamics. The reconnection rate $\langle E_z \rangle$ from our simulation is with a value as high as $\sim 0.6\ B_0V_{A,0}/c$ at the peak, much larger than the maximum rate of $0.1 - 0.2B_0V_{A,0}/c$ in anti-parallel Hall-reconnection simulations[12]. Note that the maximum rate for Hall-reconnections with guide field is generally lower than the $0.1 - 0.2B_0V_{A,0}/c$ value.

How does the drag produce the electron-scale magnetic reconnection during Buneman instability? Previously[51] we found that when $\Xi_e = 0$, the Ohm's Law (Eq. 2) can be split

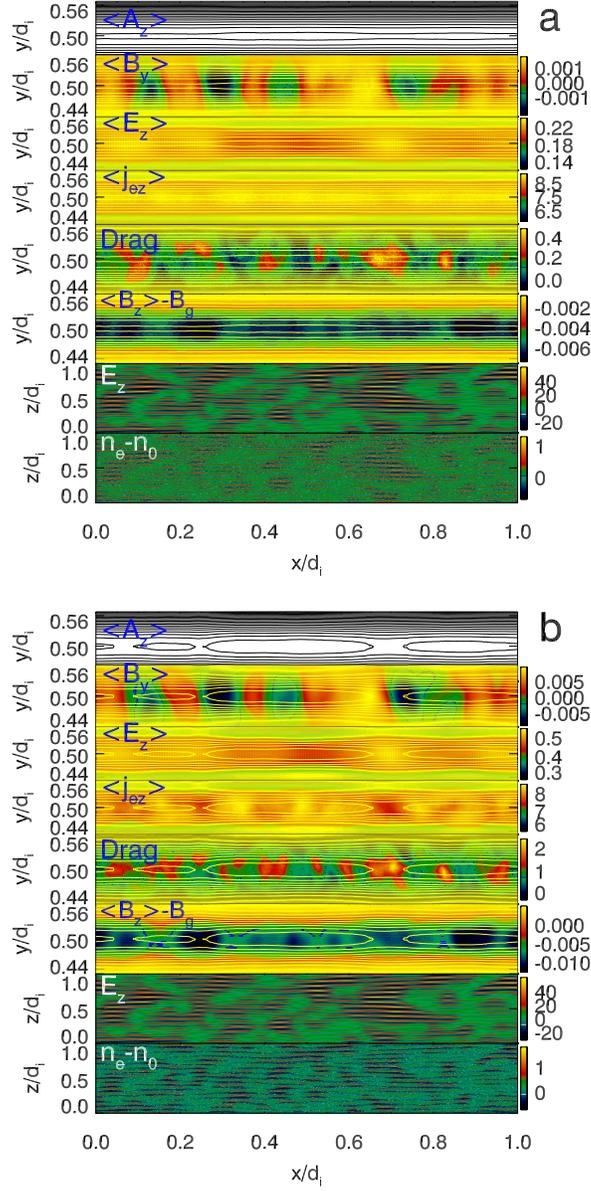

FIG. 1. (a): Quantities at $\omega_{pe,0}t = 48$ in the current sheet simulation when the Buneman instability occurs. (b): The same quantities as in (a) but at $\omega_{pe,0}t = 60$ when the Buneman instability reaches its peak. $\langle A_z \rangle$ is the z-component of the magnetic field vector potential. $E_z$ and electron density fluctuations $n_e - n_0$ are shown in the mid-plane x-z of the current sheet.

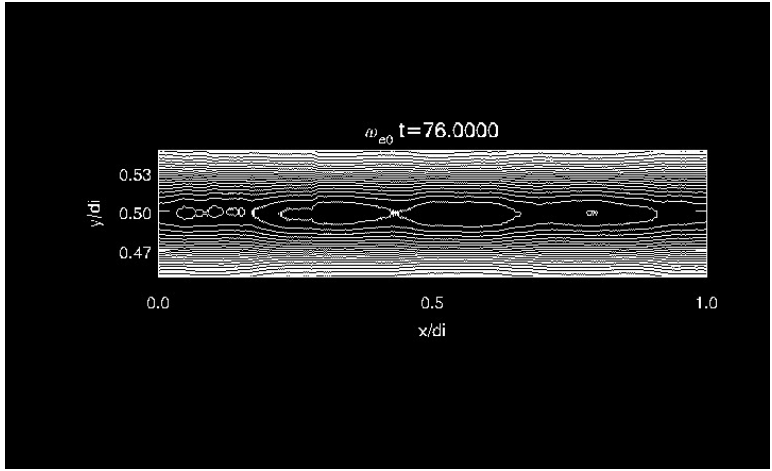

FIG. 2. The z-component of magnetic field vector potential $\langle A_z \rangle$ shows clear magnetic field line merging on the electron scale with amplitude $\langle A_z \rangle$ varying from 0 to 0.5 (Multimedia view).

into two equations around the time of the saturation of Buneman instability:

$$\langle E_z \rangle = -\frac{m_e}{e}\partial_t \langle U_{ez} \rangle + \langle D_{ez} \rangle, \tag{3}$$

$$E_z^{wv} = E_z - \langle E_z \rangle = -\frac{m_e}{e} U_{ez} \partial_z U_{ez} - \frac{1}{e\langle n_e \rangle}\partial_z P_{ezz}, \tag{4}$$

where $E_z^{wv}$ is the Buneman instability generated coherent electric field, $\langle E_z \rangle$ is the mean electric field, and $E_z = E_z^{wv} + \langle E_z \rangle$. Buneman instability produces electron holes along $z$ near the mid-plane of the current sheets. The z-averaged $\langle E_z \rangle$ cancels out local effects of the electron holes since $\langle E_z^{wv} \rangle = 0$. Eq. (3) establishes the relation between the reconnection rate and the drag while the average of Eq.(4) along z-direction indicates that the dissipation of kinetic energy by drag is converted into the parallel electron heating $\triangle \langle P_{ezz} \rangle = \triangle \langle m_e n_e U_{ez}^2 \rangle/2$ by the electron holes[71]. The inductive electric field is related to the magnetic vector potential by $\langle E_z \rangle = -\partial_t \langle A_z \rangle/c$ using the Coulomb gauge.

The wave couplings during the Buneman instability form wavepackets and cause the drag to become non-uniform within the current sheet and such non-uniformity determines the topology of the magnetic field. The spatial scale of wavepacket is determined by the uncertainty principle $\delta k \delta x \sim 2\pi$, where $\delta k$ and $\delta x$ are the width of the wave number and the spatial size of the wave packet respectively. We approximate $\delta x \sim V_g \delta t$ and $\delta t \sim \omega_{pe}^{-1}$, where $V_g \sim d\omega/dt$. Using the $\omega - k$ relation for Buneman instability, we obtain $\delta k \sim k_{fast}$, $k_{fast}$ is the wave number of the fastest growing mode and satisfies $k_{fast} = \omega_{pe}/v_d$. In Fig. 1, both the reconnection electric field and the drag appear "clumpy" in the x-direction and form "wave-packets" with scale $\sim 0.1 d_i = 0.1\sqrt{m_i/m_e} d_e = d_e$ for $m_i/m_e = 100$, close to the wavelength of Buneman instability. Combining the relation $\langle E_z \rangle = -\partial_t \langle A_z \rangle/c$ and the Ampère's law, we obtain the equation for $\langle A_z \rangle$ around the mid-plane of the current sheet:

$$\partial^2 \langle A_z \rangle - \frac{1}{c^2} \partial_t^2 \langle A_z \rangle + \frac{4\pi}{c} \langle j_{ez} \rangle = 0, \qquad (5)$$

where $\partial^2 \equiv \partial_x^2 + \partial_y^2$, and we have neglected the contribution from $j_{ex}$ which is much smaller than $j_{ez}$. Eq.5 is a standard d'Alembert equation. The near field solution is

$$\langle A_z(\mathbf{x}) \rangle = \frac{1}{c} \int \frac{\langle j_{ez}(\mathbf{x}', t - r/c) \rangle}{r} dx dy, \qquad (6)$$

where $\mathbf{r} = \mathbf{x}' - \mathbf{x}$. The magnetic topology is determined by the inhomogeneity of the current sheet which is coupled to the drag by

$$\frac{1}{c} \partial_t \langle A_z(\mathbf{x}) \rangle = \frac{m_e}{e} \partial_t \langle j_{ez}/n_e \rangle - D_{ez}. \qquad (7)$$

The magnetic vector potential is the near field solution of d'Alembert equation and the magnetic field is a quasi-static field, and its change in the xy plane describes the change of magnetic field topology $\langle B_x \rangle = -\partial_y \langle A_z \rangle$ and $\langle B_y \rangle = -\partial_x \langle A_z \rangle$. Using Eq.(7), we estimate how the electron-scale magnetic reconnection is determined by the inhomogeneity of the drag. From Eq.(7) we have $\gamma \langle A_z(\mathbf{x}) \rangle/c \sim 4\pi e \gamma \triangle \langle j_{ez} \rangle/\omega_{pe}^2 - D_{ez}$, where $\gamma \sim \omega_{pe}$ is the growth rate of Buneman instability. Then we have $\langle A_z(\mathbf{x}) \rangle \sim 4\pi e \triangle \langle j_{ez} \rangle d_e/\omega_{pe} - D_{ez} d_e$, where $d_e$ is the electron inertial length $c/\omega_{pe}$. Thus $\langle B_y \rangle \sim (4\pi e \triangle \langle j_{ez} \rangle/\omega_{pe} - D_{ez}) d_e/\triangle_x$, where $\triangle_x$ is the spatial scale of inhomogeneity of drag. It is clear that the spatial scale of the inhomogeneous turbulence drag is an important factor that determines the electron-scale magnetic reconnection.

It should be noted that the electron-scale magnetic reconnection is not necessary for the fast dissipation of magnetic energy during Buneman instability. The topology change of the field lines in Buneman instability requires the inhomogeneity of drag. The inhomogeneity is caused by wave couplings that break the symmetry of magnetic potential. To illustrate this point, we conduct a 2D Buneman instability simulation in the yz plane so that no electron-scale magnetic reconnection can develop. Fig. 3 shows the electric field $E_z$ at $\omega_{pe,0} t = 60$. The intense localized bi-polar electric fields, i.e. electron holes form along the z-direction and the drag produced by these electron holes can dissipate the kinetic energy and the associated magnetic energy. At the same time, electron heating is produced in directions parallel and perpendicular to the magnetic field. The time evolutions of electron kinetic, magnetic energy and pressure are shown in Fig. 4, which are nearly identical to the evolutions of these quantities averaged over the x-direction in the 3D Buneman simulation as we found previously[51]. It is clear that the inductive electric field is produced due to the fast dissipation of magnetic field (Fig. 5) whose time evolution is similar to that in the 3D simulation and the peak dissipation rate is $\sim 0.6B_0V_{A0}/c$. In Fig. 5, $B_y$ generated by the Buneman instability oscillates on spatial scale $\sim 0.1d_i$ similar to the wavelength of Buneman turbulence along z in the 2D simulation. The mean value of $\langle B_y \rangle$ is $\sim 10^{-6} \sim 0$, this confirms that no electron-scale magnetic reconnection develops in the 2D simulation. For comparison, the evolution of $B_y$ along z at the same time in the 3D Buneman simulation is also shown. $B_y$ obviously deviates from zero. Beside the fast oscillations with spatial scale 0.1 $d_i$ along z caused by the waves generated by the Buneman instability, a new feature is the long spatial scale variation caused by the wave-wave coupling of Buneman instability which leads to the nonzero $\langle B_y \rangle$. The value of $\langle B_y \rangle$ is 0.01 which is consistent with the value shown in Fig 1 for 3D Buneman instability. This difference clearly demonstrates that $B_y$ is caused by the inhomogeneity of the drag.

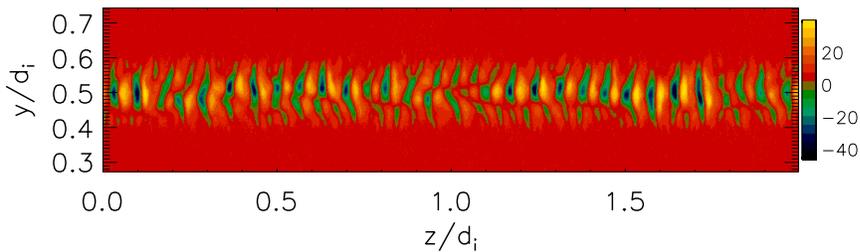

FIG. 3. $E_z/E_0$ in the 2D Buneman instability simulation at $\omega_{pe,0}t = 60$, when electron holes are developed.

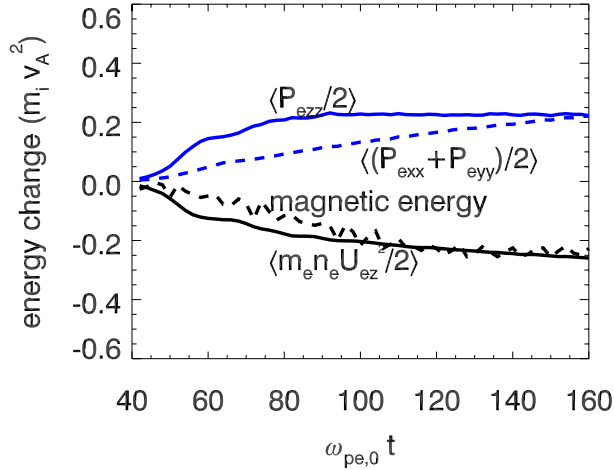

FIG. 4. The time evolution of energy change in 2D Buneman instability simulation. Black solid: kinetic energy of the electron beam; black dashed line: magnetic energy. Blue solid line: electron parallel thermal energy; blue dashed line: perpendicular thermal energy.

To summarize, AR has two different but connected effects. First, it dissipates the magnetic energy and produces a mean inductive electric field; Second, the inhomogeneity of drag causes inhomogeneous magnetic energy dissipation which breaks the local symmetry of the magnetic field potential in the current sheet and leads to electron-scale magnetic reconnection. The electron-scale reconnection rate is significantly higher than Hall-reconnection rate. With this newly gained insight of Bumenan instability induced AR, we can look into the role of AR driven in ion-scale magnetic reconnection.

## C. How does Anomalous Resistivity Accelerate Magnetic Reconnection?

Since the Buneman instability grows along the current sheet perpendicular to the reconnection plane, the instability cannot develop in 2D reconnection simulations. Comparing

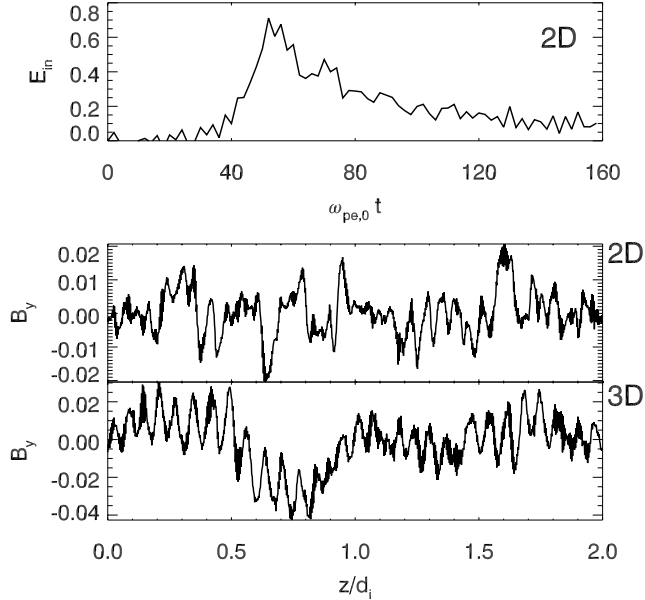

FIG. 5. Top panel: The time evolution of inductive electric field $\langle E_z \rangle$ in the 2D Buneman instability simulation. Bottom panel: The $B_y$ generated along z at $\omega_{pe,0} = 60$ in the 2D and 3D Buneman simulations.

our 3D and 2D simulations allows clear demonstration of the effects of turbulence generated by the Buneman instability. As we have shown previously, in the 3D magnetic reconnection simulation, turbulence driven by an ES Buneman instability and an EM electron velocity shear instability makes the magnetic reconnection significantly faster than the non-turbulent 2D reconnection[37]. The Buneman instability produces electron holes in the early stage for a brief period around $\Omega_{i0}t \sim 0.4$ ($\omega_{pe,0}t = 80$) and the electron velocity shear instability starts to widen the current sheet and entails the filamentary structures at later stage. Close examination of Fig.4 in Che et al. (2011)[37] reveals that before the onset of Buneman instability, the reconnection electric field $\langle E_z \rangle$ at the x-line (reconnection is in the xy plane) is fully supported by the electron inertia. As the Buneman instability is triggered at $\Omega_{i0}t = 3$ ($\omega_{pe,0}t = 600$) when the electron drift is about $6V_{A0} \sim 2v_{te}$, the increasing drag reduces the electron acceleration, and eventually the drag becomes comparable to the electron inertia at $\Omega_{i0}t = 3.2$ ($\omega_{pe,0}t = 640$) and lasts to $\Omega_{i0}t = 3.4$ ($\omega_{pe,0}t = 680$). During this period the

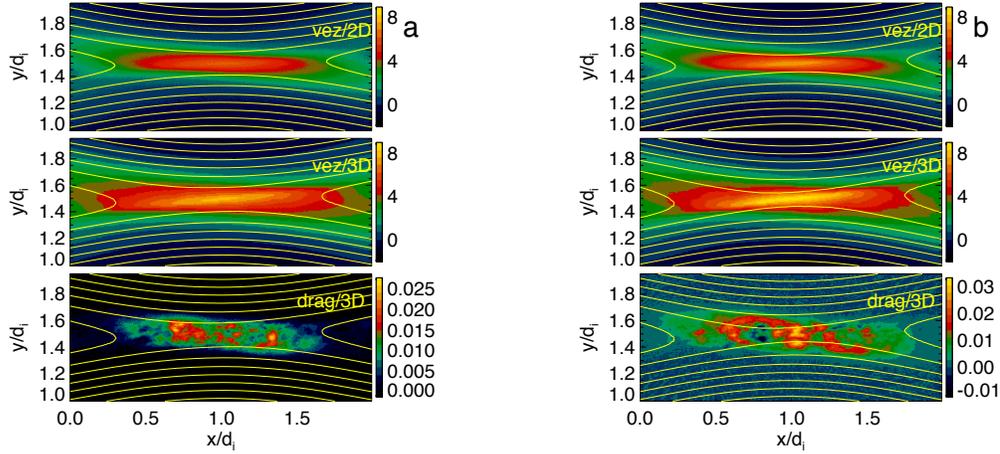

FIG. 6. The top two panels in **a** & **b** show the magnitude of electron velocity $V_{ez}$ in 3D and 2D magnetic reconnection, and the bottom panels show the drag $D_{ez}$. **a**: the onset of Buneman instability at $\Omega_{i0}t = 3$ ($\omega_{pe,0}t = 600$); **b**: the saturation of Buneman instability at $\Omega_{i0}t = 3.2$ ($\omega_{pe,0}t = 640$). The solid lines represent the contour of magnetic vector potential $\langle A_z \rangle$.

Buneman instability becomes saturated. Then the Buneman instability starts to decay while the electron velocity shear instability starts to grow and surpasses the Buneman instability at $\Omega_{i0}t \sim 3.8$. Then the induced whistler wave turbulence becomes dominant. After that the repeating occurrences of electron velocity shear instability continuously maintain the fast reconnection.

To understand how drag/AR accelerate magnetic reconnection, we need to answer two separate questions: 1) How does the drag break magnetic field lines around the x-point? and 2) can the drag affect the dynamics in the IDR?

In the preceding sections we have shown that AR can dissipate magnetic energy, produce inductive electric field and break magnetic field lines by its inhomogeneity in electron-scale thin current sheet. Here we show that the field-lines are also broken due to the inhomogeneity of the drag in the vicinity of the x-line to accelerate magnetic reconnection.

In Fig. 6 we show the electron velocity $V_{ez}$ in the 2D reconnection simulation and $\langle V_{ez} \rangle$

in the 3D reconnection simulation at $\Omega_{i0}t = 3$ ($\omega_{pe,0}t = 600$) when Buneman instability is triggered, and at $\Omega_{i0}t = 3.2$ ($\omega_{pe,0}t = 640$) when it reaches its saturation. The saturation lasts to $\Omega_{i0}t = 3.4$ ($\omega_{pe,0}t = 680$) when the drag becomes comparable to the inertia as we have shown previously[37]. The electron velocity $V_{ez}$ peaks around the x-point and exceeds the threshold $v_{te}$, triggering Buneman instability. As a result, intensive drag is produced in the thin current layer around the x-point. We show the drag $D_{ez}$ at $\Omega_{i0}t = 3, 3.2$ in the xy plane in Fig. 6. Clearly the drag is spatially clumpy. While the drag is produced by Buneman instability at the vicinity of the x-point, the plasma outflow brings the inhomogeneous drag away from the x-point into the IDR, as illustrated by the expansion of the turbulent current sheet in Fig. 6. It should be noted that the AR induced dissipation reduces the acceleration of electron beams, but the electron beams are still continuously accelerated by the enhanced reconnection electric field and maintain the continuous growth of Buneman instability. Both the width and length of the current layer in the reconnection plane are larger in the 3D than in the 2D simulation, implying that the plasma heating produced by AR (drag) could play a similar role as collisional resistivity.

As the drag breaks the field lines at the x-point, the non-zero $B_y$ with opposite sign at the two sides of the x-point is produced, causing the null-point to shift in a stochastic manner in the 3D magnetic reconnection. In Fig.7, we show $E_z$ at $\Omega_{i0}t = 3.2$ in the yz plane at $x = d_i$, i.e. the location of the x-point in the non-turbulent 2D reconnection. Bi-polar localized intense electric fields or electron holes form along z due to the Buneman instability near the x-point. In the bottom panel, we show $B_y(z)$ at the fiducial 2D x-point at times of $\Omega_{i0}t = 2.5$ (blue line) prior to the onset of the Buneman instability, and $\Omega_{i0}t = 3.2$ (black line) when the Buneman instability is near the peak. Large deviation of $B_y$ from zero is similar to $B_y$ produced in the 3D Buneman instability simulation shown in Fig.5. The deviation of $B_y$ from zero averaged over z is about 0.01 at $\Omega_{i0}t = 3.2$ while the mean value of $B_y$ is approximately zero at $\Omega_{i0}t = 2.5$. This stochastic shift of the x-point causes the EDR to become broader.

We further examine how the Buneman instability enhances the reconnection electric field $\langle E_z \rangle$. Fig. 8 shows $\langle E_z \rangle$ in the xy plane at $\Omega_i t = 3.2$ ($\omega_{pe,0}t = 640$) in both the 3D and 2D simulations. Significant differences are found: 1) The reconnection electric field in the 3D simulation is concentrated in a region centered at the x-point and extends in the x-direction. On the other hand the reconnection electric field is more diffused in the 2D simulation, indicating that in 3D the reconnection rate is much higher in the vicinity of the x-point as well as in the outflow where the drag is carried out than in regions where AR is small. Away from the x-point the electric field is similar to the mean field in the 2D simulation; 2) In the 3D simulation, the profile of $\langle E_z \rangle$ at the x-point resembles a "wavepacket". The spatial scale $\sim 0.1 - 0.2 d_i$ is consistent with the Buneman "wavepackets" in the current sheet simulation (c.f. Fig. 1) — implying that the reconnection is mediated by the inhomogeneity of the drag in the thin current sheet, and electron-scale magnetic reconnections occur around the x-line.

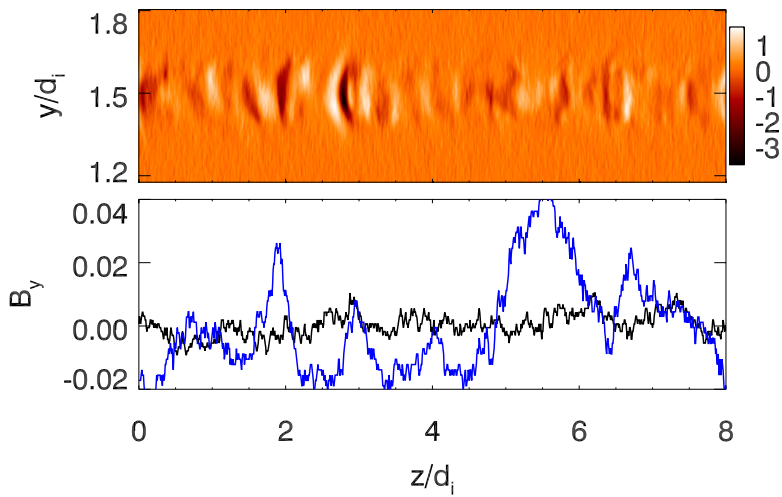

FIG. 7. Top panel: $E_z$ in yz plane at $\Omega_{i0}t = 3.2$ in 3D magnetic reconnection simulation. Bottom Panel: The evolution of $B_y$ along z at $\Omega_{i0}t = 2.5$ (black line) prior to the trigger of Buneman instability and $\Omega_{i0}t = 3.2$ (blue line) near the peak of Buneman instability in 3D magnetic reconnection simulation.

An important feature of the electron-scale magnetic reconnection is the enhancement of $\langle B_y \rangle$ by the inhomogeneous drag (Fig.6). We show $\langle B_y \rangle$ at $\Omega_{i0}t = 3.2$ in both 2D and 3D magnetic reconnections in Fig. 9. $\langle B_y \rangle$ is increased by $\sim 0.01$ in the area extending from the EDR to the IDR where $B_y^{2D} \sim 0.04$ (taken at $x = 1.6d_i$ and $y = 1.5$ in the 2D reconnection), or $\sim 25\%$ enhancement in 3D than in 2D. The enhancement of $\langle B_y \rangle$ is asymmetric around the x-point—this is due to the Hall-effect which generates an asymmetric quadratic electron current density in guide field magnetic reconnection[19,30].

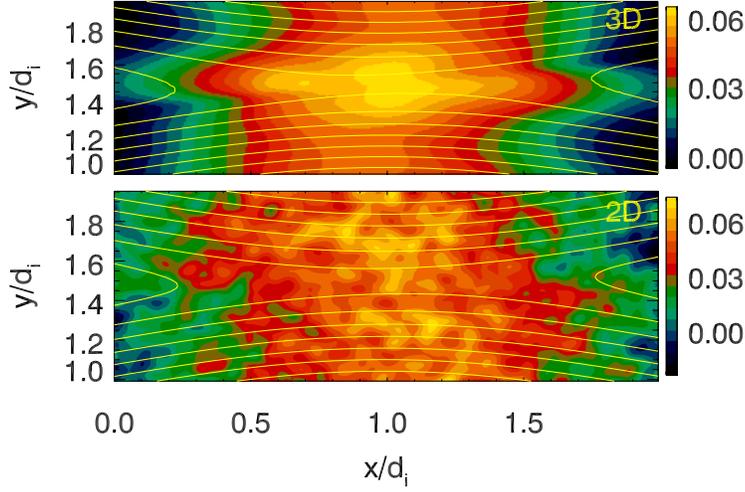

FIG. 8. The reconnection electric field $\langle E_z \rangle$ generated in 3D and $E_z$ in 2D magnetic reconnection at the saturation of Buneman instability $\Omega_i t = 3.2$ ($\omega_{pe,0} t = 640$). The solid lines are the corresponding magnetic field lines.

It can be shown that the enhancement of $B_y$ (denoted as $\langle B_y^{bun} \rangle$) in the IDR is related to the drag at the x-point in a rather simple way. In §II B we found $\langle B_y^{bun} \rangle \sim (4\pi e \triangle \langle j_{ez} \rangle / \omega_{pe} - D_{ez}) d_e / \triangle_x$. The out-of-plane current $j_{ez}$ in the reconnection region is maintained by the reconnection electric field and the variation along x is small as seen in the simulation. On the other hand, the time for the plasma outflow to travel to the IDR is much shorter than the electron hole decay time, thus the drag is carried along with the outflow into the IDR. Neglecting the small decrease of $\langle j_{ez} \rangle$ we have $\langle B_y^{bun} \rangle \sim D_{ez} d_e / \triangle_x$. From Fig.6, $D_{ez} \sim 0.02$ and $\triangle_x \sim d_e$, we obtain $\langle B_y^{bun} \rangle \sim D_{ez} \sim 0.02$, which is consistent with the value in Fig.9.

The enhanced $B_y$ can increase the opening angle of local field lines at the x-line. However, in our simulation the increase of the angle $\triangle \theta$ in IDR is very small which can be estimated by $\triangle \theta \sim \langle B_y^{bun} \rangle \tan \theta / \tan \theta' \sim \frac{1}{2} \langle B_y^{bun} \rangle \sin(2\theta) <\sim \frac{1}{2} \langle B_y^{bun} \rangle \sim 0.01$ (about $0.5°$), where

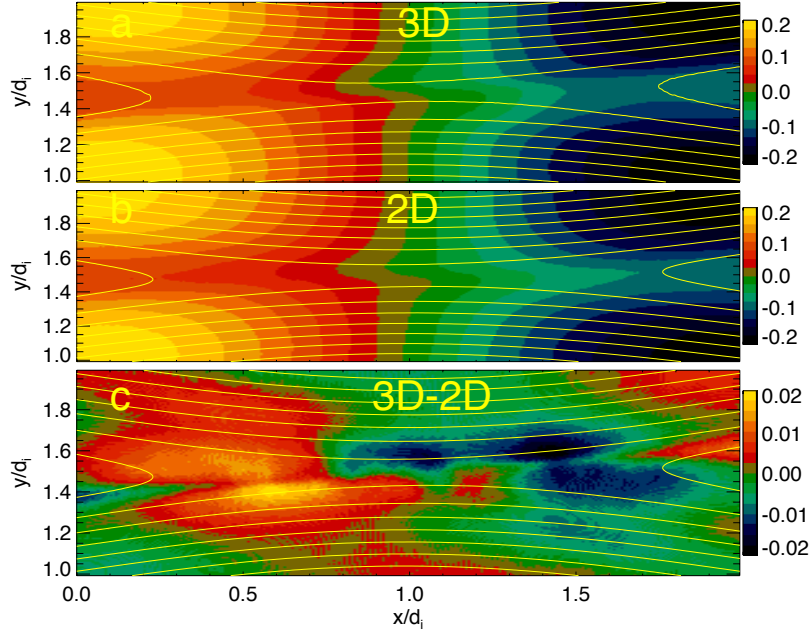

FIG. 9. $\langle B_y \rangle$ generated in 3D and $B_y$ in 2D magnetic reconnection at the peak of Buneman instability $\Omega_{i0}t = 3.2$ ($\omega_{pe,0}t = 640$). In bottom panel, the difference between $\langle B_y \rangle$ and $B_y$ is displayed. The contour of magnetic vector potential $\langle A_z \rangle$ is shown as solid lines.

$\tan\theta = B_y/B_x$.

We now look at how the AR generated dissipation affect the magnetic reconnection. In Fig.10, we study the Poynting equation $\partial w/\partial t + \nabla \cdot \mathbf{S} + \mathbf{j}_e \cdot \mathbf{E} = 0$, where $w = B^2/4\pi$ is the magnetic energy density. We consider the electromagnetic energy changes over the time from $\Omega_{i0}t = 3$ to $\Omega_{i0}t = 3.2$ ($\triangle t = 0.2\Omega_{i0}^{-1}$), the period when the Buneman instability grows to saturation. Since $j_{ez}E_z$ contributes most to the heating $\mathbf{j}_e \cdot \mathbf{E}$, we neglect the x and y components. We can see that $j_{ez}^{2D}E_z^{2D}$ in the EDR of 2D reconnection is $\sim 0.5$ while $\langle j_{ez}E_z \rangle$ in the EDR of 3D reconnection is $\sim 0.8$. In 2D $j_{ez}^{2D}E_z^{2D}$ corresponds to the magnetic energy that is converted into the kinetic energy of the electron beams near the x-line due to the reconnection electric field acceleration. In 3D $\langle j_{ez}E_z \rangle$ includes two parts according to Eq.(3) and (4), i.e. $\langle j_{ez}\rangle\langle E_z\rangle$ and $\langle j_{ez}E_z^{wv}\rangle$. The term $\langle j_{ez}E_z^{wv}\rangle = \langle \delta j_{ez}E_z^{wv}\rangle$ is the electron heating produced by electron holes. Given that $\langle j_{ez}\rangle\langle E_z\rangle = -m_e\langle j_{ez}\rangle\partial_t\langle U_{ez}\rangle/e + \langle j_{ez}\rangle D_{ez}$, the term $\langle j_{ez}\rangle\langle E_z\rangle$ includes the electron anomalous heating and the gain of the kinetic energy of electron beams by acceleration. In Fig. 6, the velocities of the electron beams in both the 2D and 3D reconnection are roughly the same, meaning the same amount of magnetic energy is used to accelerate the electron beams in both cases, and therefore the AR induced electron heating can be estimated by $\langle j_{ez}E_z\rangle - j_{ez}^{2D}E_z^{2D}$ shown in Fig. 10 A, and $\sim 30-50\%$ of the magnetic energy turns into the thermal energy of electrons in the AR dominated phase of the reconnection.

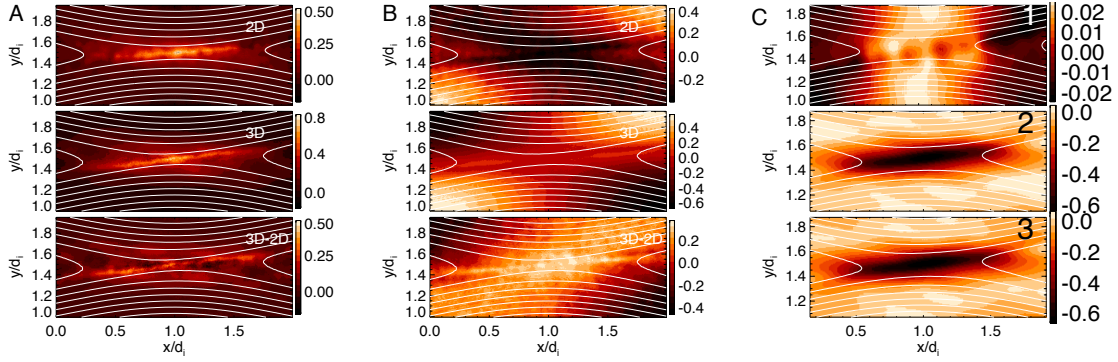

FIG. 10. Terms in the Poynting equation in the magnetic reconnection simulations. **A**. top panels: $\langle j_{ez}E_z\rangle$ in the 3D simulation and $j_{ez}^{2D}E_z^{2D}$ in the 2D simulation at $\Omega_{i0}t = 3.2$; bottom panel: $\langle j_{ez}E_z\rangle - j_{ez}^{2D}E_z^{2D}$. **B**. Top panels: the change of the magnetic energy density $\triangle(B_x^2+B_y^2+B_z^2)/8\pi$ in the 2D and $\triangle\langle(B_x^2+B_y^2+B_z^2)/8\pi\rangle$ in the 3D simulations between $\Omega_{i0}t = 3$ and $\Omega_{i0}t = 3.2$, divided by $0.2\Omega_{i0}^{-1}$. These are rough estimates of the average rate of change of the magnetic energy $\partial w/\partial t$; the bottom panel: The difference between 3D and 2D rates. **C**: The divergence of Poynting flux. 1) $\partial_x S_x$, 2) $\partial_y S_y$, 3) $\partial_x S_x + \partial_y S_y$.

We show the change of magnetic energy in Fig. 10B. The rate of magnetic energy loss $\partial w/\partial t$ near the x-line in the 2D reconnection approximately matches $j_{ez}^{2D}E_z^{2D}$. However, in the 3D reconnection, $\partial w/\partial t$ in the EDR is close to zero—indicating that there should be an extra Poynting flux flowing into the EDR. The difference between the magnetic energy change rates in the 3D and 2D reconnections further illustrates that magnetic energy is flowing into the EDR in the 3D reconnection. Obviously the Buneman instability at the x-point is responsible for such energy transfer. To illustrate the magnetic energy transfer, we examine the divergence of Poynting flux. Consistent with what we see in the 2D simulation $\partial_t w + j_{ez}^{2D}E_z^{2D} \sim 0$, the divergence of Poynting flux in the EDR is close to zero (not shown here). The divergence of Poynting flux in the 3D reconnection is shown in Fig. 10C. We found the Poynting flux flowing into EDR in y direction $(c/4\pi)\partial_y\langle E_z B_x\rangle$ (panel C2) balances the dissipation which accounts for $\sim 40\%$ of the released magnetic energy in the EDR, in other words the AR enhanced $E_z$ is responsible for the magnetic energy transfer into the EDR to accelerate the reconnection. Compared to the in-flow of magnetic energy, $\sim 10\%$ of the released magnetic energy is brought out to the IDR by the enhanced $B_y$ through the x-component of the Poynting flux $\sim -(c/4\pi)\partial_y\langle E_z B_y\rangle$ (panel C1).

## III. DISCUSSION AND CONCLUSIONS

In this paper, we explore how Buneman instability induced AR/drag at the x-line accelerates 3D guide-field magnetic reconnection. With a 3D PIC simulation of Buneman instability in an electron scale current sheet we show that 1) drag $D_{ez}$ can dissipate magnetic energy stored in the current layer $j_{ez}$ through the dissipation of the kinetic energy of electron beams. This process is described by the Ohm's law $\langle E_z\rangle = -\frac{m_e}{e}\partial_t\langle j_{ez}/n_e\rangle + D_{ez}$; 2) the coupling of waves leads to the formation of wavepackets, causing the drag $D_{ez}$ to become inhomogeneous along x. As a result, the inductive electric field $\langle E_z\rangle$ is also a spatial function, which in turn leads to fast impulsive electron-scale magnetic reconnections as the inhomogeneous drag breaks the symmetry of the magnetic potential through $-\partial_t\langle A_z\rangle/c = \langle E_z\rangle$, and generates the perpendicular magnetic field $\langle B_y^{bun}\rangle = -\partial_x\langle A_z\rangle$.

The electron-scale magnetic reconnection caused by AR is not affected by Hall-effect. The reconnection rate can reach as high as 0.6 $B_0 V_{A0}/c$, and such a high rate has not been seen previously in simulations. The electron-scale magnetic reconnection however does not alter the overall energy conversion rates between kinetic, thermal and magnetic energy in the current sheet, as shown in a comparison between 2D and 3D Buneman instability simulations.

In ion-scale reconnection simulations, we demonstrate how AR affects magnetic reconnection. We were able to separate the effects of AR from other non-turbulent effects that accelerate magnetic reconnection by comparing the full 3D reconnection simulation with a benchmark 2D simulation in which Buneman instability does not develop. It should be noted that in the 3D ion-scale reconnection simulation, the Buneman instability developed at the x-line is weaker than that in the electron-scale current sheet simulation. This is because 1) the initial electron drift in reconnection simulation is $4V_{A0}$ and the instability is triggered as the drift increased to 6 $V_{A0} \sim 2v_{te}$, much smaller than the initial electron drift $9V_{A0} \sim 3v_{te}$ in the current sheet simulation; 2) while the velocity of electron beams eventually reaches 9 $V_{A0}$ at later times of the simulation, the parallel electron thermal velocity also increases to 5 $V_{A0}$ due to the electron heating produced by the Buneman instability. As a result the electron holes and the inductive electric field produced in the 3D reconnection simulation are all much weaker than that in the Buneman instability simulation even though the 3D magnetic reconnection is still much faster than the corresponding 2D magnetic reconnection.

With these simulations, we have found the following: 1) The drag significantly increase the reconnection electric field at the x-point and in the reconnection out-flow region extending to the IDR; 2) The inhomogeneity of the drag can break the magnetic field lines to produce electron-scale magnetic reconnections at the x-line; 3) The drag can be brought out by the reconnection out-flow into the IDR, thus increases the size of the EDR; 4) The drag also increases the size of the EDR by stochastic shifts of the x-point; 5) The enhanced perpendicular magnetic field $B_y$ is increased by an average of $\sim 25\%$ in our simulation; and 6) About 40% of the released magnetic energy is converted into the electron thermal energy by AR while 50% is converted into the kinetic energy of the electron beams through the acceleration by reconnection electric field. The enhanced magnetic energy dissipation is supported by a net Poynting flux in-flow. About 10% of the released magnetic energy is brought out by the enhanced Poynting flux out-flow.

We have shown that $B_y^{bun}$ is related to the drag by $B_y^{bun} \sim D_{ez}d_e/\triangle_x$, where $\triangle_x$ is the spatial scale of inhomogeneity of drag. Therefore only small $\triangle_x$ can lead to large $B_y^{bun}$ (in our case $\triangle_x \sim d_e$) — implying perhaps only drag with electron scale inhomogeneity has the potential to significantly accelerate magnetic reconnection, probably faster than the Hall reconnection rate of 0.1 $V_A$. This may explain why some PIC reconnection simulations with electron holes developed do not show one-to-one correspondence between reconnection rate and energy conversion[73–75].

A group of electron beam instabilities commonly known as streaming instabilities which include Buneman instability, electron two-stream instability[60] and lower hybrid instability[62] etc, can all develop around the x-line and in the separatrix as the reconnection current sheets proceed to the electron inertial length. These electron electrostatic instabilities can form coherent structures such as electron holes and efficiently dissipate both of the kinetic and magnetic energy via AR. The results presented in this paper serve as a case study which should be useful for more general inquiries into the role of AR in magnetic reconnection.

The results presented in this paper are potentially observable in the magnetosphere. The MMS time resolution of $\sim \omega_{pe}^{-1}$ for high frequency waves[76] can resolve the waves generated during the nonlinear phase of the Buneman instability, which lasts for $\omega_{pe,0}t \sim 80$. Interesting enough, a spontaneous small-scale fast guide field reconnection produced by the twisted magnetic flux tube may have been observed by MMS[77]. Streaming instabilities and electron holes in EDRs of magnetic reconnections are also discovered by MMS[32,54–58]. With more observed events, the MMS mission is in a good position to observationally determine the role of turbulence in magnetic reconnection[78,79].

## ACKNOWLEDGMENTS

This research is supported by the NASA *MMS* in association with NASA contract NNG04EB99C. The author thanks the entire *MMS* project for the support, in particular the discussions with colleagues in the FPI team: J. Dorelli, W. Paterson, B. Giles, M. Goldstein, L. Avanov, B. Lavraud, M. Chandler, D. Gershman and C. Schiff, and the participants of 2016 1st MMS community Science Workshop in UCLA. The simulations and analysis were carried out at the NASA Advanced Supercomputing facility at the Ames Research Center and the National Energy Research Scientific Computing Center. The author thanks the anonymous referee for the critical and constructive comments that helps to improve the clarity of this manuscript.